\documentclass{aa}  

\usepackage{graphicx}
\usepackage{txfonts}
\begin{document}

   \title{Tomography of silicate dust around M-type AGB stars}

   \subtitle{I. Diagnostics based on dynamical models}

   \author{S. Bladh
          \inst{1}
          \and
          C. Paladini\inst{2}
          \and
          S. H\"ofner\inst{3}
          \and 
          B. Aringer\inst{1}
                    }

   \institute{Dipartimento di Fisica e Astronomia Galileo Galilei, Università di Padova, Vicolo dell’Osservatorio 3, 35122 Padova, Italy\label{inst1}\\
              \email{sara.bladh@unipd.it}
         	\and
	    Institut d'Astronomie et d'Astrophysique, Universit\'e Libre de Bruxelles, CP. 226, Boulevard du Triomphe, 1050 Brussels, Belgium \label{inst2}
             \and 
            Department of Physics and Astronomy, Uppsala University, Box 516, 75120 Uppsala, Sweden\label{inst3}
             }

   \date{Received; accepted}

 
  \abstract
   {The heavy mass loss observed in evolved asymptotic giant branch stars is usually attributed to a two-step process: atmospheric levitation by pulsation-induced shock waves, followed by radiative acceleration of newly formed dust grains. Detailed wind models suggest that the outflows of M-type AGB stars may be triggered by photon scattering on Fe-free silicates with grain sizes of about 0.1 - 1~$\mu$m. As a consequence of the low grain temperature, these Fe-free silicates can condense close to the star, but they do not produce the characteristic mid-IR features that are often observed in M-type AGB stars. However, it is probable that the silicate grains are gradually enriched with Fe as they move away from the star, to a degree where the grain temperature stays below the sublimation temperature, but is high enough to produce emission features.}
   {We investigate whether differences in grain temperature in the inner wind region, which are related to changes in the grain composition, can be detected with current interferometric techniques, in order to put constraints on the wind mechanism.}
   {We use phase-dependent radial structures of the atmosphere and wind of an M-type AGB star, produced with the 1D radiation-hydrodynamical code DARWIN, to investigate if current interferometric techniques can differentiate between the temperature structures that give rise to the same overall spectral energy distribution.}
   {The spectral energy distribution is found to be a poor indicator of different temperature profiles and therefore is not a good tool for distinguishing different scenarios of changing grain composition. However, spatially resolved interferometric observations have promising potential. They show signatures even for Fe-free silicates (found at 2-3 stellar radii), in contrast to the spectral energy distribution. Observations with baselines that probe spatial scales of about 4 stellar radii and beyond are suitable for tracing changes in grain composition, since this is where effects of Fe enrichment should be found.}
   {}


   \maketitle
%

\section{Introduction}
The massive winds of evolved asymptotic giant branch (AGB) stars are commonly assumed to be driven by radiation pressure on dust grains, which form in the outer atmospheric layers of these cool luminous giants. Pulsations and large-scale convective motions trigger atmospheric shock waves, resulting in extended dynamical atmospheres with cool and dense mass layers that provide favorable conditions for dust formation and grain growth. Photons emitted by the star transfer momentum to the newly formed dust particles by absorption or scattering, and the dust grains accelerate the surrounding gas through collisions, leading to an outflow if the radiative acceleration overcomes the gravitational attraction of the star. 

The scenario of radiation pressure on dust as a wind driver dates back to the the first detections of characteristic circumstellar emission features at mid-IR wavelengths \citep[e.g.][]{wickramasinghe1966,gehrz1971,sedlmayr1994,habing2004}. Prominent examples are the 10 and 18~$\mu$m features in M-type AGB stars. These features have been identified as belonging to magnesium-iron silicates, such as olivine ([Mg,Fe]$_2$SiO$_4$) and pyroxene ([Mg,Fe]SiO$_3$), and are caused by stretching and bending resonances in the SiO$_4$-tetrahedron \citep[e.g.][]{wolf1969,low1970}. These resonance features can be attributed to both amorphous and crystalline silicate grains, but the broadness of the observed 10 and 18~$\mu$m features in evolved stars indicate that they are predominantly formed by amorphous silicates \citep{molster2002a,molster2002b}.

Detailed dynamical atmosphere and wind models suggest that the outflows of M-type AGB stars may be triggered by photon scattering on virtually Fe-free magnesium silicates with grain sizes of about 0.1 - 1~$\mu$m \citep{hoefner2008,bladh2015}, which is supported by recent discoveries of dust particles in this size range in the close vicinity of several cool giants \citep[e.g.][]{norris2012,ohnaka2016,ohnaka2017}. Fe-free silicate particles are very transparent at near-IR wavelengths, around the stellar flux maximum, which leads to low radiative heating and, consequently, to low grain temperatures. While this favors condensation close to the star, it means that the Fe-free silicates are too cool to produce significant mid-IR emission features \citep[see, e.g., the discussion in][]{bladh2015}. However, it is likely that the grains are gradually enriched with Fe as they move away from the star, as more opaque materials become stable against radiative heating with increasing distance and decreasing radiative flux.  A self-regulating process probably keeps the Fe content at a level where the grain temperature is just below the sublimation temperature throughout the dust condensation zone \citep[i.e., the Fe/Mg ratio of the silicates acts as a thermostat;][]{woitke2006}. 

In this paper we investigate whether such a change in grain composition over the first few stellar radii in the outflow produces signatures that can be detected with spatially resolved mid-IR observations. Based on structures obtained with detailed wind models, we simulate the effects of changing grain composition on mid-IR visibilities for a range of baselines, tracing the variation of the 10~$\mu$m silicate feature with distance from the star. We also test whether different enrichment scenarios, which results in different temperature structures, can be distinguished with data from past and present spectro-interferometric instrumentation when the mid-IR spectra of the objects are basically identical in unresolved observations. 

There is an extensive archive of interferometric data for M-type AGB stars available to the scientific community, consisting of observations taken by the mid-IR instrument of the Very Large Telescope Interferometer \citep[VLTI/MIDI, decommissioned in March 2015,][]{leinert2003}. So far, just a few of these observations have been compared with synthetic observables from detailed models of the atmosphere and wind. A first attempt to compare interferometric mid-IR observations with detailed wind models for M-type AGB stars was made by \cite{sacuto2013}, confirming the presence of silicate grains in the close stellar environment of RT Vir. The present paper should also be viewed in light of the upcoming second-generation VLTI thermal and mid-IR imager MATISSE \citep{lopez2006} that will be installed in Paranal in January 2018. This facility will allow again the possibility of using interferometric mid-IR data to probe the wind acceleration zone and constrain the wind mechanism in AGB stars. 

The current theoretical effort is a demonstration of principle, exploring the diagnostic potential of mid-IR interferometry regarding grain composition close to the star in the context of dynamical atmosphere and wind models. It will be followed by a series of papers comparing various models with existing spatially resolved interferometric observations for M-type AGB stars \citep[e.g., data acquired within the VLTI/MIDI AGB Large Program;][]{paladini2017}.

The paper is organized in the following way. In Sect.~\ref{darwin} we describe the modeling method (the dynamical atmosphere and wind models, the a posteriori radiative transfer calculations of synthetic observables, and the approaches to simulating changing grain composition). In Sect.~\ref{res} we investigate if we can use interferometric techniques to differentiate between temperature profiles of grain enrichment scenarios that produce similar spectral energy distributions. Finally, in Sect.~\ref{conclusion} we summarize our results. 


\section{Modeling methods}
\label{darwin}
\subsection{Dynamical structures}
The radial structures of the extended atmosphere and wind formation regions are modeled using the 1D radiation-hydrodynamical code DARWIN \citep[for a detailed description see][]{hoefner2016}. The models cover a spherical shell with an inner boundary situated just below the photosphere and an outer boundary in accordance with the dynamical properties of the model (at about 20-30 stellar radii for models that develop a wind). The variable structure of the atmosphere is produced by simultaneously solving the equations of hydrodynamics (equation of continuity, equation of motion and the energy equation), the frequency-dependent radiative transfer (describing the momentum and energy balance of the radiation field) and the time-dependent description for grain growth of magnesium silicates (Mg$_2$SiO$_4$). 

The DARWIN models are defined by the input parameters of stellar mass $M_*$, stellar luminosity $L_*$, and effective temperature $T_*$ of the star. The variability of the star is simulated by a sinusoidal variation of velocity and luminosity at the inner boundary of the model, characterized by a pulsation period $P$ and the amplitudes of the velocity and luminosity variations (described by the parameters $u_\textrm{p}$ and $f_\textrm{L}$). Since dust nucleation (i.e., the initial formation of condensation nuclei from the gas phase) is still poorly understood for stars with C/O < 1 \citep[see, e.g.,][and reference therein]{gobrecht2016,gail2016}, we assume that tiny seed particles (with sizes corresponding to 1000 monomers, i.e., the basic building blocks of the solid) exist prior to the onset of grain growth, and we parameterize them by their abundance relative to hydrogen (described by the parameter $n_{\textrm{gr}}/n_\textrm{H}$). These seed particles will start to grow when the thermodynamic conditions are favorable. The Mg$_2$SiO$_4$ grains are assumed to be spherical and the optical properties are calculated from data by \cite{jaeger2003}, using Mie theory. The radiative acceleration of the dust depends on the dominant direction of photon scattering which may be more favorable for grains of more complex shapes. Assuming that the dust particles are spherical probably corresponds to a lower limit.

\subsection{Synthetic observables: a posteriori radiative transfer}
The DARWIN code produces time-series of snapshots of the radial structure of the atmosphere and wind. Each snapshot provides information about properties such as velocity, temperature, density, and dust grain size as a function of radial distance and time. Selected snapshots are further processed in an a posteriori radiative transfer calculation in order to produce opacity sampling spectra with a resolution of $R=20\,000$, covering the wavelength range between 0.1 and 25~$\mu$m, using the opacities from the COMA code \citep{aringer2016}. For both the DARWIN models and the a posteriori spectral and interferometric calculations, we assume solar abundances from \cite{grevesse1989} except for C, N, and O where we took the data from \cite{grevesse1994}, and the treatment of gas and dust is consistent. The optical data for Mg$_2$SiO$_4$ and MgFeSiO$_4$ is taken from \cite{jaeger2003} and \cite{dor1995}, respectively, and the optical properties of the core-mantle grains are calculated with the routine BHCOAT taken from \cite{bohren1983}.

Another output of the radiative transfer are the spatial intensity profiles for every frequency point in the simulations. The intensity profiles are then convolved with a series of rectangular filter curves covering the N band (8-13~$\mu$m) with spectral resolution $R\sim35$. The spectral resolution is chosen to match that of the low resolution mode available for the VLTI/MIDI instruments. The visibility profiles are calculated with the Hankel transform of the intensities following the approach described in \citep{paladini2009}. For the purpose of this work we assumed the model star to be at a distance of 100~pc, and we selected the typical projected baselines of 10, 30, 50, and 70~m available on the current VLTI facility.

\subsection{Simulating changing grain composition}
\label{gcomp}
The Mg/Fe ratio of magnesium-iron silicates strongly influences their absorption properties in the near-IR and visual wavelength regions where AGB stars emit most of their stellar flux. Models by \cite{woitke2006} indicate that once Fe inclusion is possible, it will control the temperature of the dust particles and act as a thermostat, keeping the grain temperature just below the sublimation limit. Consequently, the radial temperature profile of silicates is related to the grain composition (the ratio of Mg/Fe) at a given distance from the star. The question is whether that information can be used to deduce at what distance from the star Fe atoms can condense onto the silicate grains and how the grain composition changes in the wind formation region. To investigate this we simulate the effects of Fe inclusions in the a posteriori spectral calculations (i) by directly changing the temperature profile of the silicates or (ii) by altering the optical properties of the silicate grains and then calculating the grain temperature. 

\begin{figure}
\centering
\includegraphics[width=\linewidth]{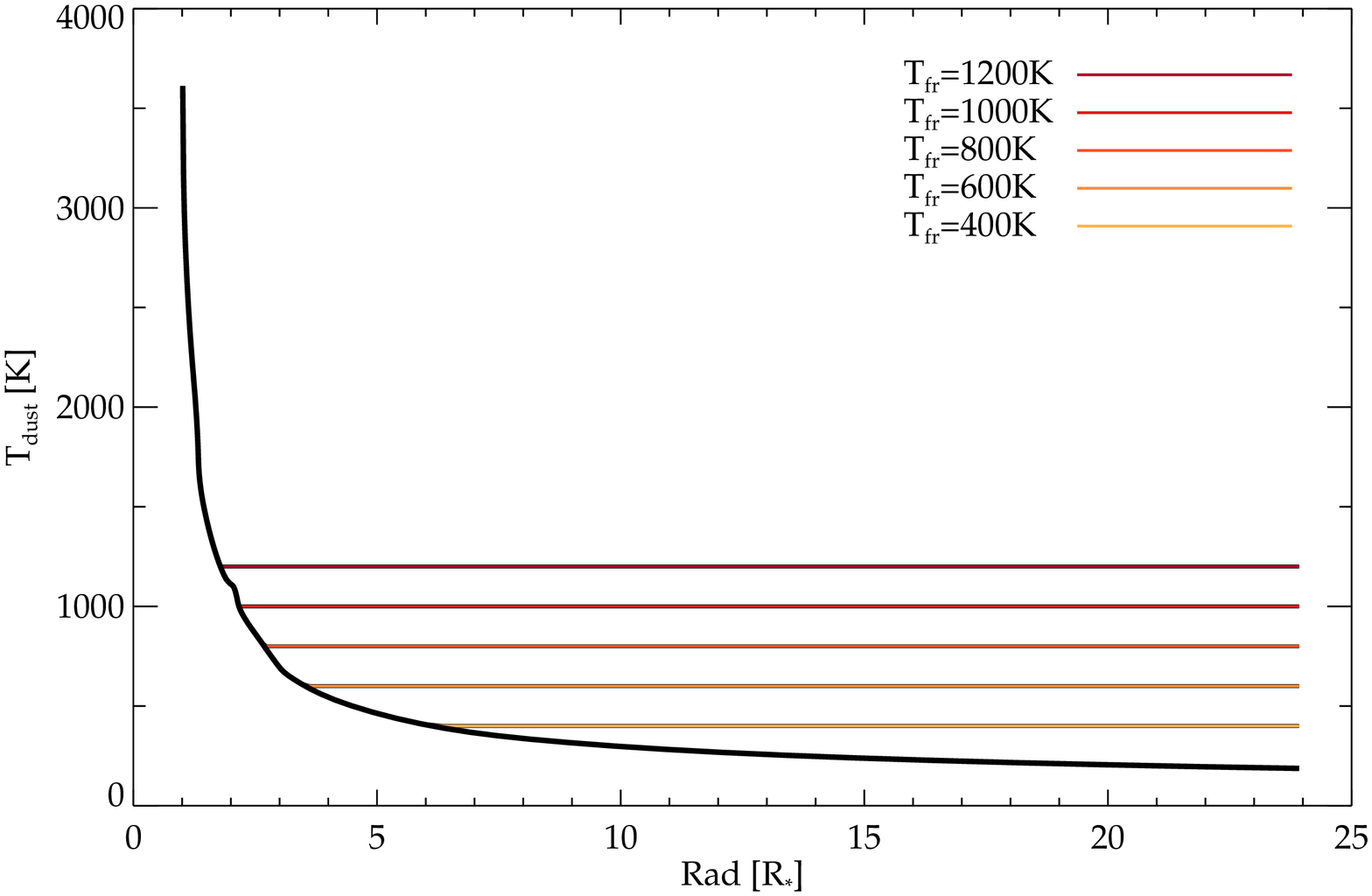}
\includegraphics[width=\linewidth]{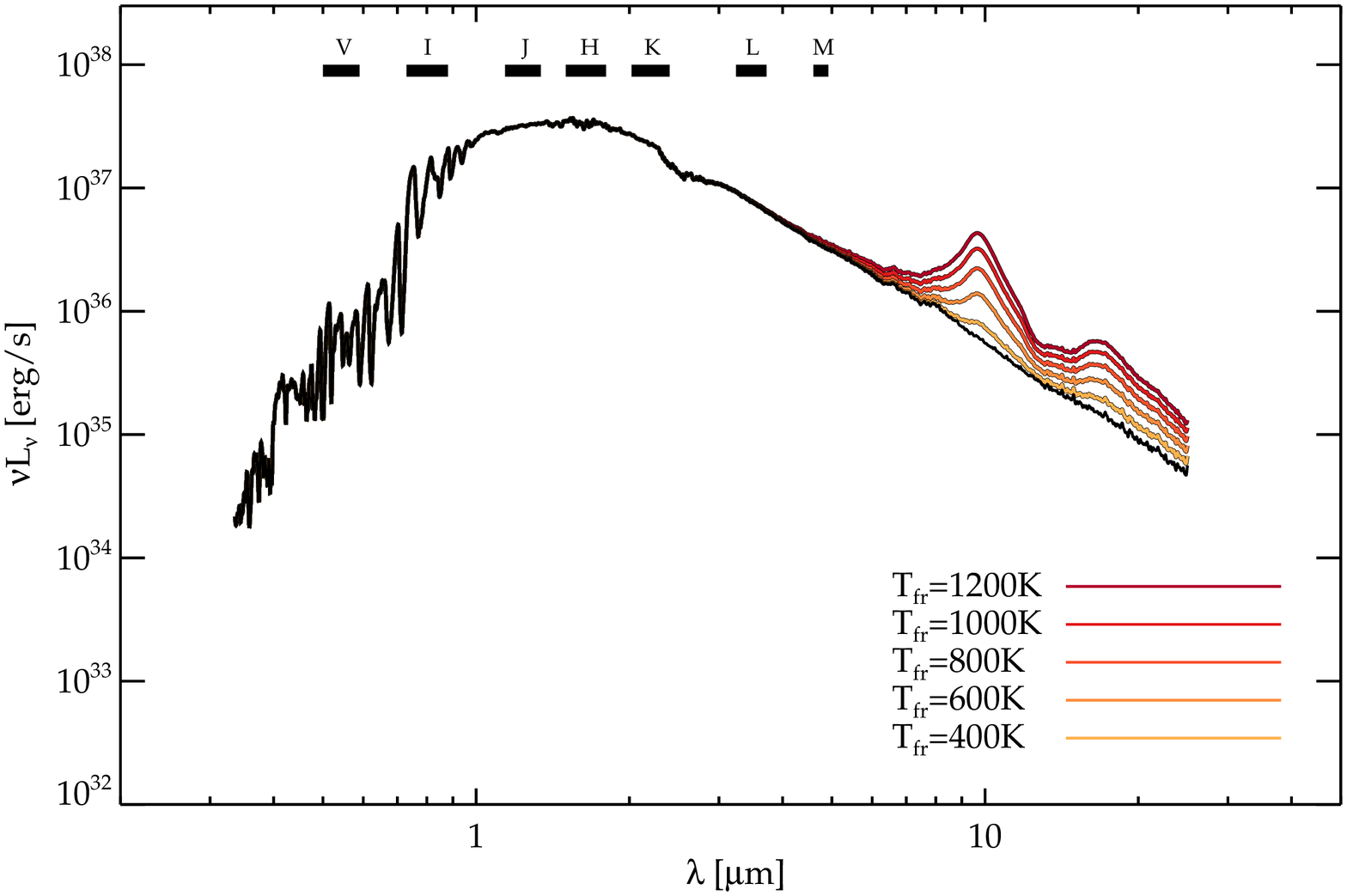}
   \caption{\textit{Top panel:} Grain temperature as a function of distance from the star. \textit{Lower panel:} Spectral energy distributions as a function of wavelength. Structure from model A3 during the maximum phase, with the minimum grain temperature $T_{\mathrm{fr}}$ set to 400K, 600K, 800K, 1000K and 1200K (the black curve shows the original model structure).}
    \label{f_freezing}
\end{figure}

As a result of the low absorption cross section of pure magnesium silicates, the grain temperature in the current DARWIN models for M-type AGB stars decreases rapidly with distance from the star. This can be seen in the upper panel of Fig.~\ref{f_freezing}, which shows the radial dependence of the grain temperature (black curve) in a typical model. To simulate the effects of Fe inclusions we set a lower limit to the grain temperature, $T_{\mathrm{fr}}$, and keep the temperature constant when it reaches this limit value, like a thermostat. Examples of such temperature profiles and the resulting spectra with different lower temperature limits are shown in  Fig.~\ref{f_freezing}. These spectra confirm that the lack of mid-IR silicate features for Fe-free grains is caused by a temperature structure that is decreasing too fast and not by a lack of dust material. However, this approach is very simplistic and does not take into account the decreasing grain temperature in the outer parts of the stellar wind.

\begin{figure}
\centering
\includegraphics[width=\linewidth]{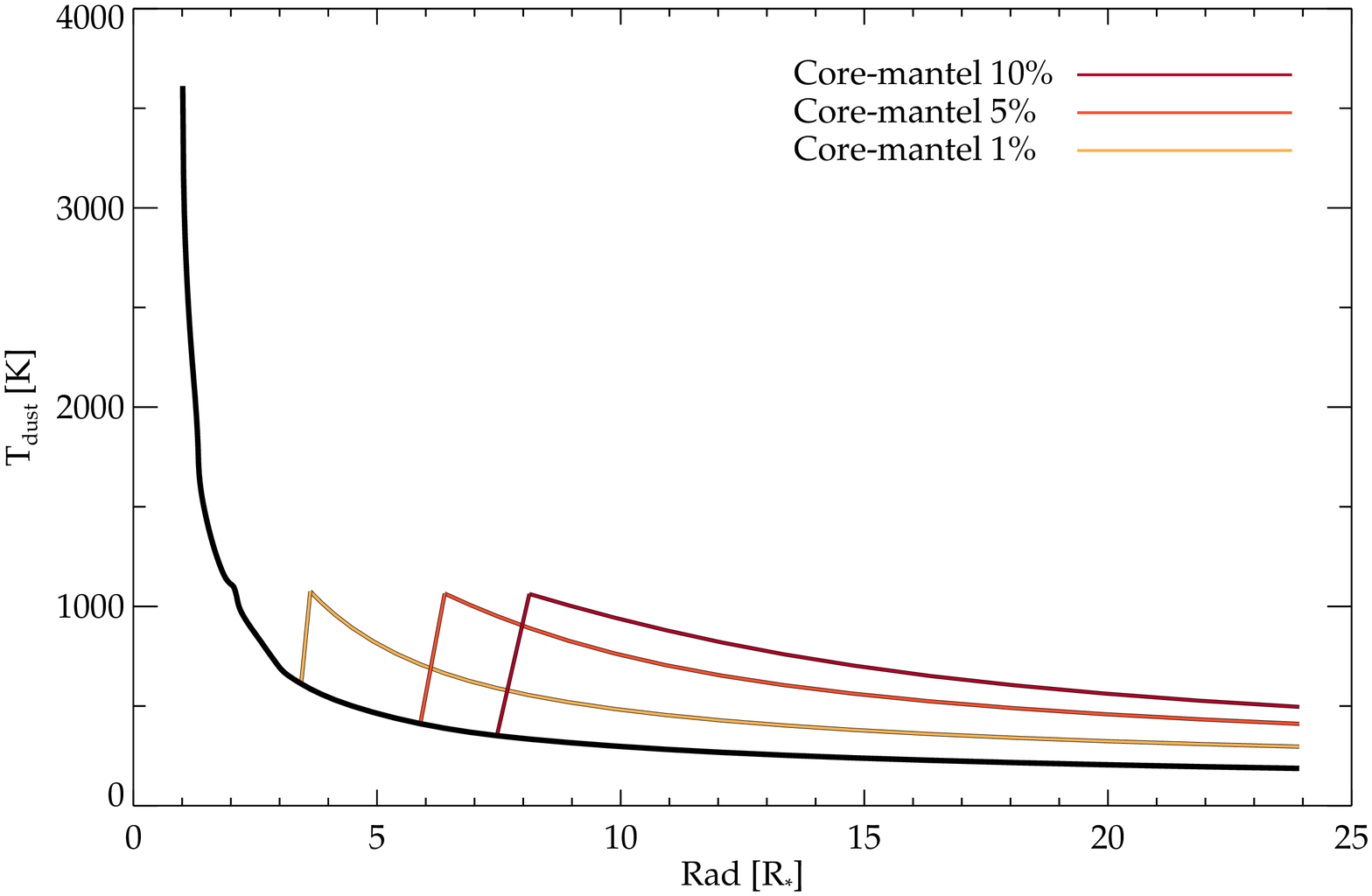}
\includegraphics[width=\linewidth]{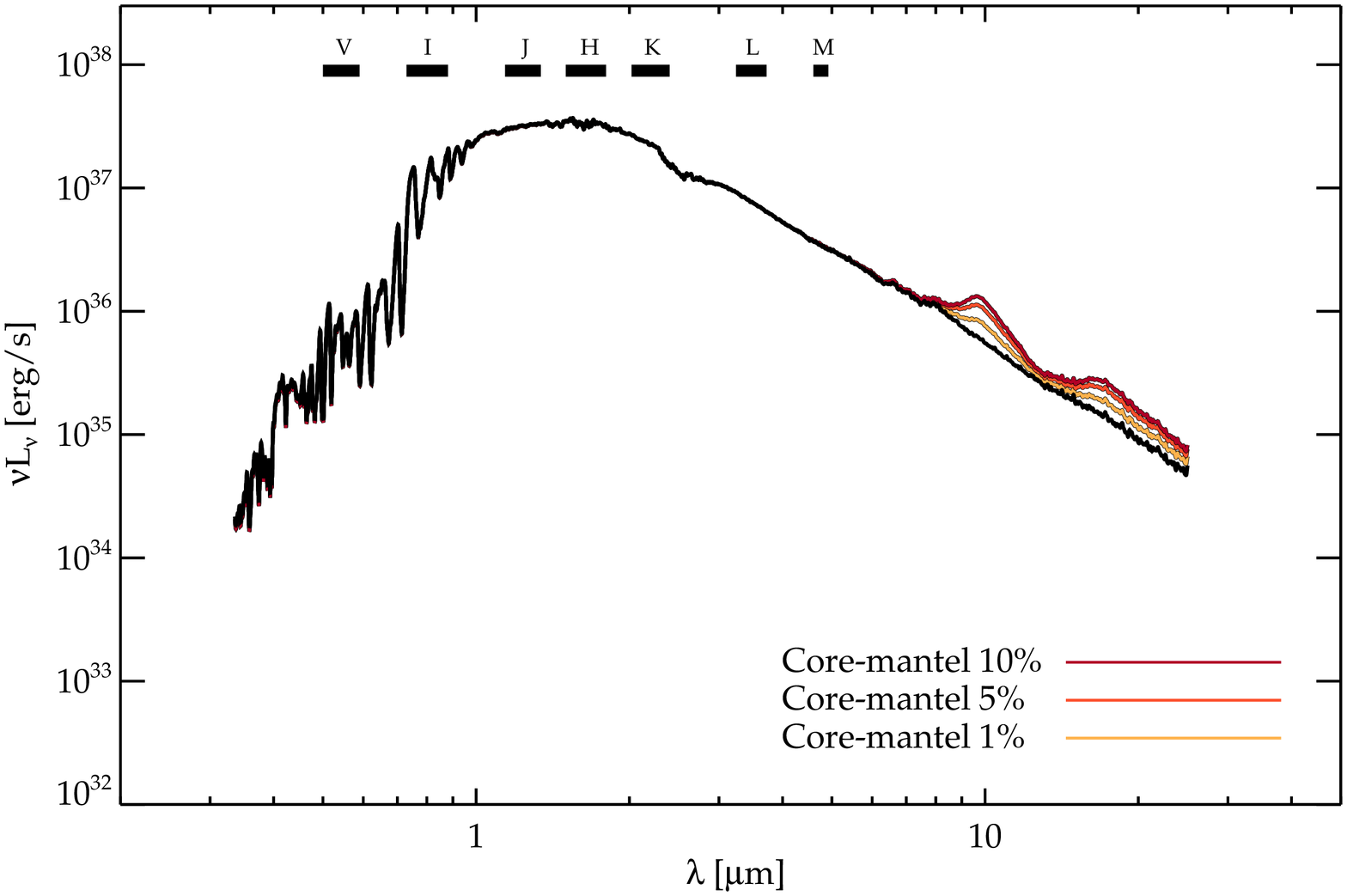}
   \caption{\textit{Top panel:} Grain temperature as a function of distance from the star. \textit{Lower panel:} Spectral energy distributions as a function of wavelength. Structure from model A3 during the maximum phase with core-mantel grains of different thicknesses (1\%, 5\% and 10\% of r$_{\mathrm{grain}}$) forming at a radial distance where they are thermally stable (the black curve shows the original model structure).}
    \label{f_coremantel}
\end{figure}

Another way to produce silicate features is to alter the optical properties by considering dust particles with a core of magnesium silicate (Mg$_2$SiO$_4$) and a thin mantle of magnesium-iron silicate (MgFeSiO$_4$), as if a small fraction of Fe atoms had condensed onto the surface of pure magnesium silicate grains. Such core-mantel grains will not be thermally stable inside the wind formation zone \citep[situated at about 2-4~R$_*$, see][]{bladh2015}, unless the Fe-bearing layer is extremely thin. We therefore use a typical model structure, but replace the optical properties and the resulting temperature profile of the pure magnesium grains (Mg$_2$SiO$_4$) with the corresponding properties for the core-mantle grains from the point where they are thermally stable. We assume this to be at the distance from the star where the temperature of the core-mantle grains drops below 1000~K \citep[approximately the temperature where magnesium-iron silicates will condense, see, e.g.,][]{gail1999}. We also assume that these grains have the same size as the magnesium silicates in the original model, but with an outer layer replaced by MgFeSiO$_4$. The resulting temperature structures for three different mantle sizes (1\%, 5\%, and 10\% of the radius, corresponding to 3\%, 15\%, and 30\% of the volume) and the corresponding spectra are shown in Fig.~\ref{f_coremantel}. As can be seen in the top panel this small addition of Fe increases the grain temperature by about 500~K even if the amount of condensed silicates remains the same.

In both approaches we assume that the atmospheric model structure will not be altered significantly by the change in grain composition and/or temperature. This is a reasonable assumption, as can be seen in the overall spectral energy distributions in the lower panels of Figs.~\ref{f_freezing} and \ref{f_coremantel}. The spectra and photometry at visual and near-IR wavelengths are basically unaffected, and only the mid-IR features change dramatically. As most of the flux from the stellar photosphere is emitted at visual and near-IR wavelengths, the dynamical impact on the atmospheric structure is dominated by the radiation pressure that is generated in the near-IR region. 

\begin{figure}
\centering
\includegraphics[width=\linewidth]{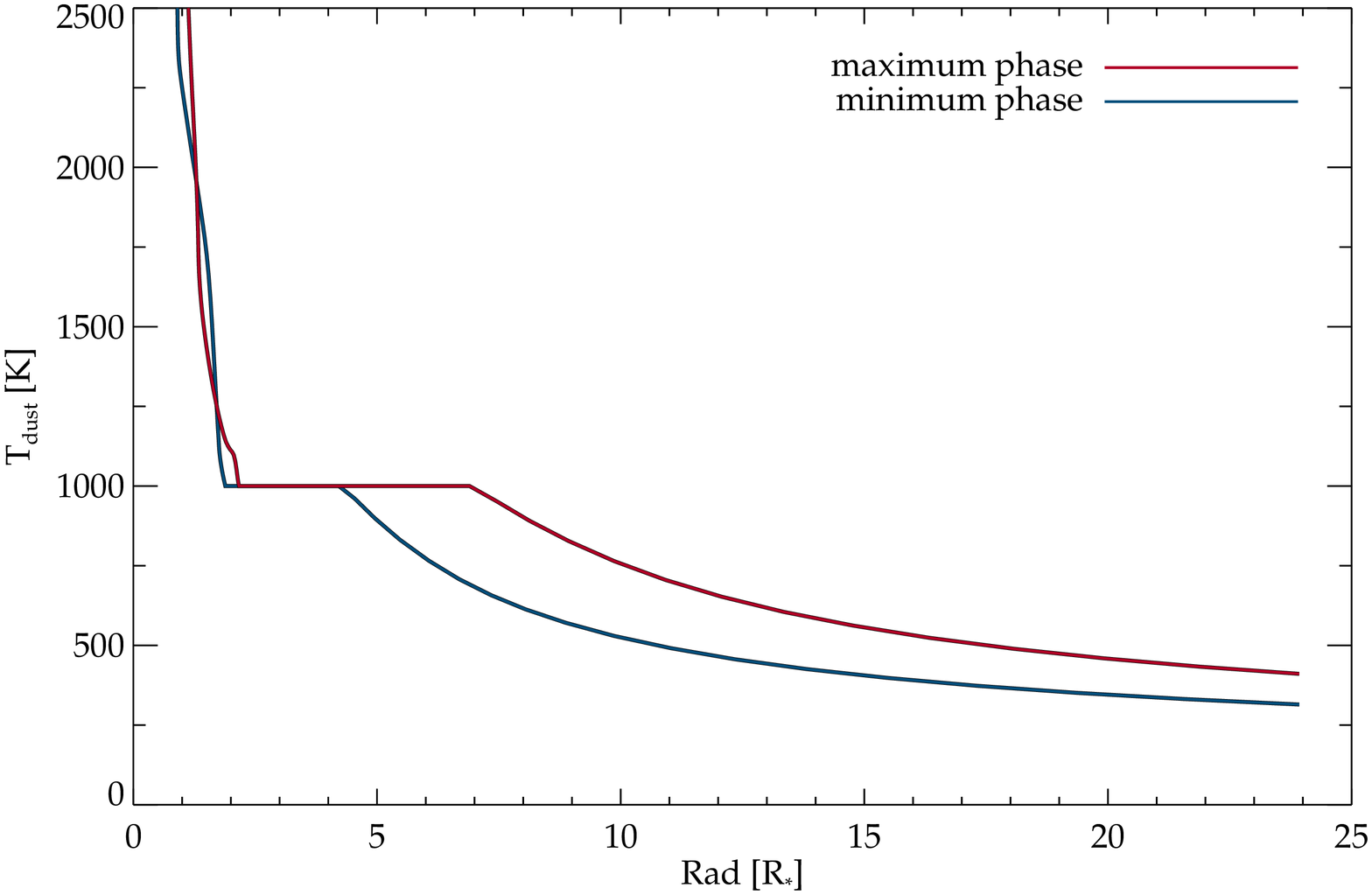}
\includegraphics[width=\linewidth]{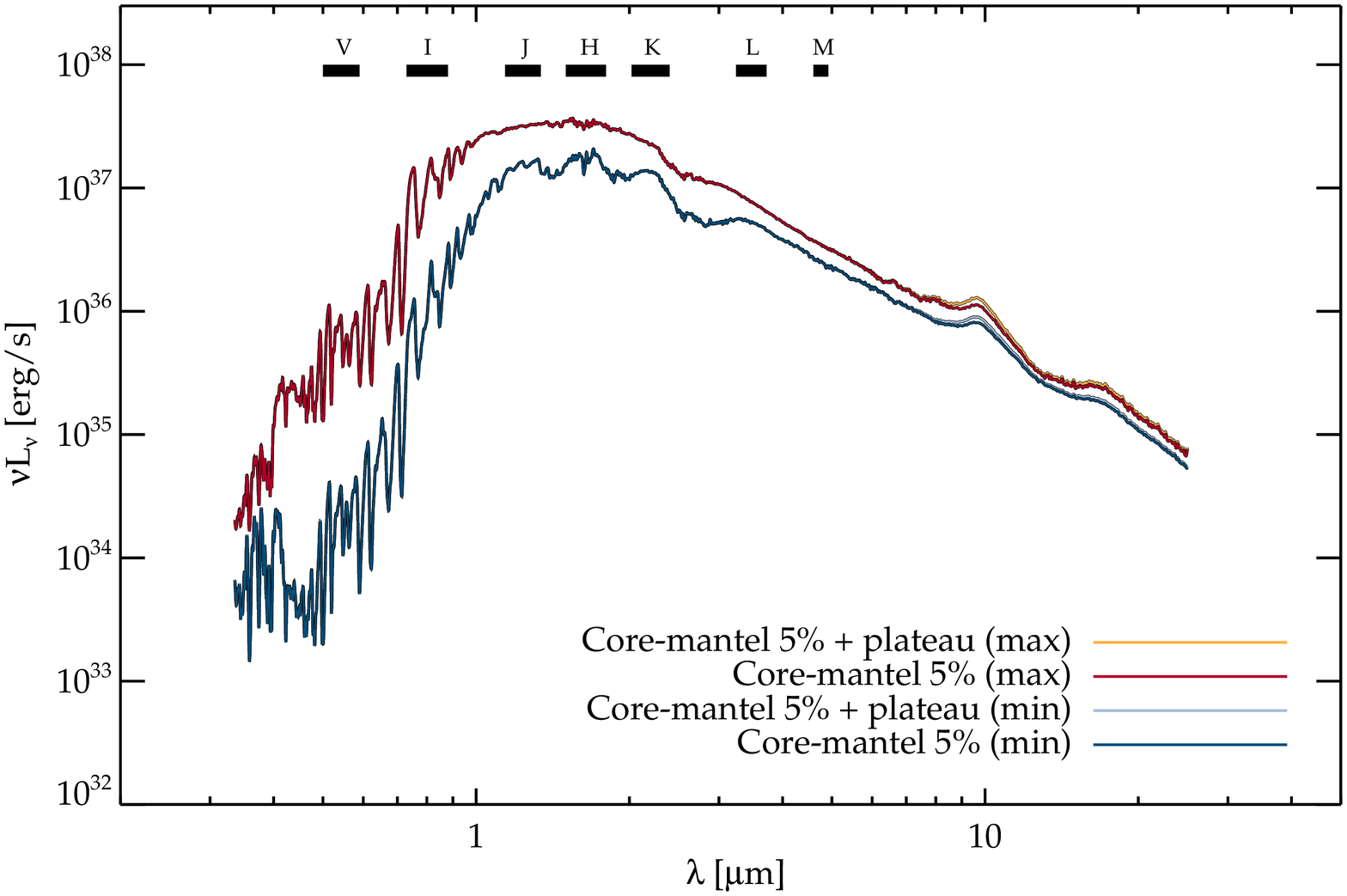}
   \caption{\textit{Top panel:} Grain temperature as a function of distance from the star. \textit{Lower panel:} Spectral energy distributions as a function of wavelength. Structures from model A3 during the maximum (red curves) and minimum (blue curves) luminosity phase. The light red and blue curves show temperature and spectra for snapshots of the atmosphere where we set a lower temperature limit of 1000\,K until the magnesium-iron layer is thermally stable.}
    \label{f_plateau}
\end{figure}

\subsection{Additional test and discussion}
The temperature profiles produced by replacing magnesium silicate grains in model A3 with core-mantle grains show a sharp increase in the grain temperature at the stability limit (see the top panel of Fig.~\ref{f_coremantel}). A more realistic approach would allow the layer of magnesium-iron silicate to grow in thickness with distance from the star as more Fe atoms can condense onto the dust particles. Since Fe inclusions act as a thermostat, keeping the grain temperature at approximately at 1000\,K, we can simulate mantle growth by setting a lower temperature limit of 1000\,K until the 5\% mantle is thermally stable. The temperature structures and spectra resulting from this construct are shown in Fig.~\ref{f_plateau}. This temperature profile differs from setting a constant lower limit to the temperature as it allows the grain temperature to drop when the density in the stellar wind decreases and dust condensation slows down. The plateau in the dust temperature before the magnesium-iron layer becomes thermally stable at a thickness of 5\% does not affect the strength of the silicate features much. Instead, the strength of the mid-IR silicate features is mostly determined by the grain temperature in the outer part of the wind acceleration zone ($R>4-5\,R_*$). This is particularly interesting as some interferometric studies seem to indicate that silicates in M-type AGB stars form at about this distance \citep[e.g.,][]{karovicova13}.

The results from \cite{karovicova13} also indicate that Al$_2$O$_3$ forms in some stars at distances of about 2 stellar radii or less, prior to silicate condensation. Al$_2$O$_3$ has been discussed as a possible alternative to silicates as a source of the scattered light observed close to several AGB stars \citep[e.g.][]{ireland2005,norris2012,ohnaka2016}, and as potential seed particles for the condensation of silicates \citep[e.g.][]{kozasa1997a,kozasa1997b}. At solar abundances, the radiation pressure due to Al$_2$O$_3$ is too low to drive a wind, but if silicates form as mantles on Al2O3 cores, this may speed up grain growth to sizes relevant for wind driving considerably. Experimental dynamical models based on such core-mantle grains tend to show higher wind velocities and mass-loss rates than models where the outflow is driven by pure silicate grains \citep[see][]{hoefner2016}. Mid-IR Al$_2$O$_3$ features are usually found in stars with a low mass-loss rate while silicate features dominate for higher mass-loss rates \citep{lorenz2000}.

\section{Results}
\label{res}

In order to investigate the observable signatures of changing Fe content in silicate grains in the inner wind region, we use model A3 from \cite{bladh2013}. This wind model has stellar parameters typical of an M-type AGB star ($M=1\,M_{\odot}$, $L=5000\,L_{\odot}$, $T_*=2800\,K$); pulsation parameters $P=310\,d$, $u_{\mathrm{p}}=4$\,km/s, and $f_L=2$; and a seed particle abundance of \mbox{$n_{\mathrm{d}}/n_{\mathrm{H}}=3\times 10^{-15}$}. It produces temporal variations in visual and near-IR colors in good agrement with observations \citep[see][]{bladh2013}. The average mass-loss rate and wind velocity are $8\times 10^{-7}\,M_{\odot}$/yr and 10\,km/s, respectively, and the average grain size and degree of condensed silicon in the outer layers are $a_{\mathrm{gr}}=0.36~\mu$m and $f_{\mathrm{Si}}=0.22$. The a posteriori radiative transfer calculations are carried out using snapshots of the radial structure during the maximum and minimum luminosity phase.

\subsection{Spectral energy distribution}
\label{spec}
\begin{figure}
\centering
\includegraphics[width=\linewidth]{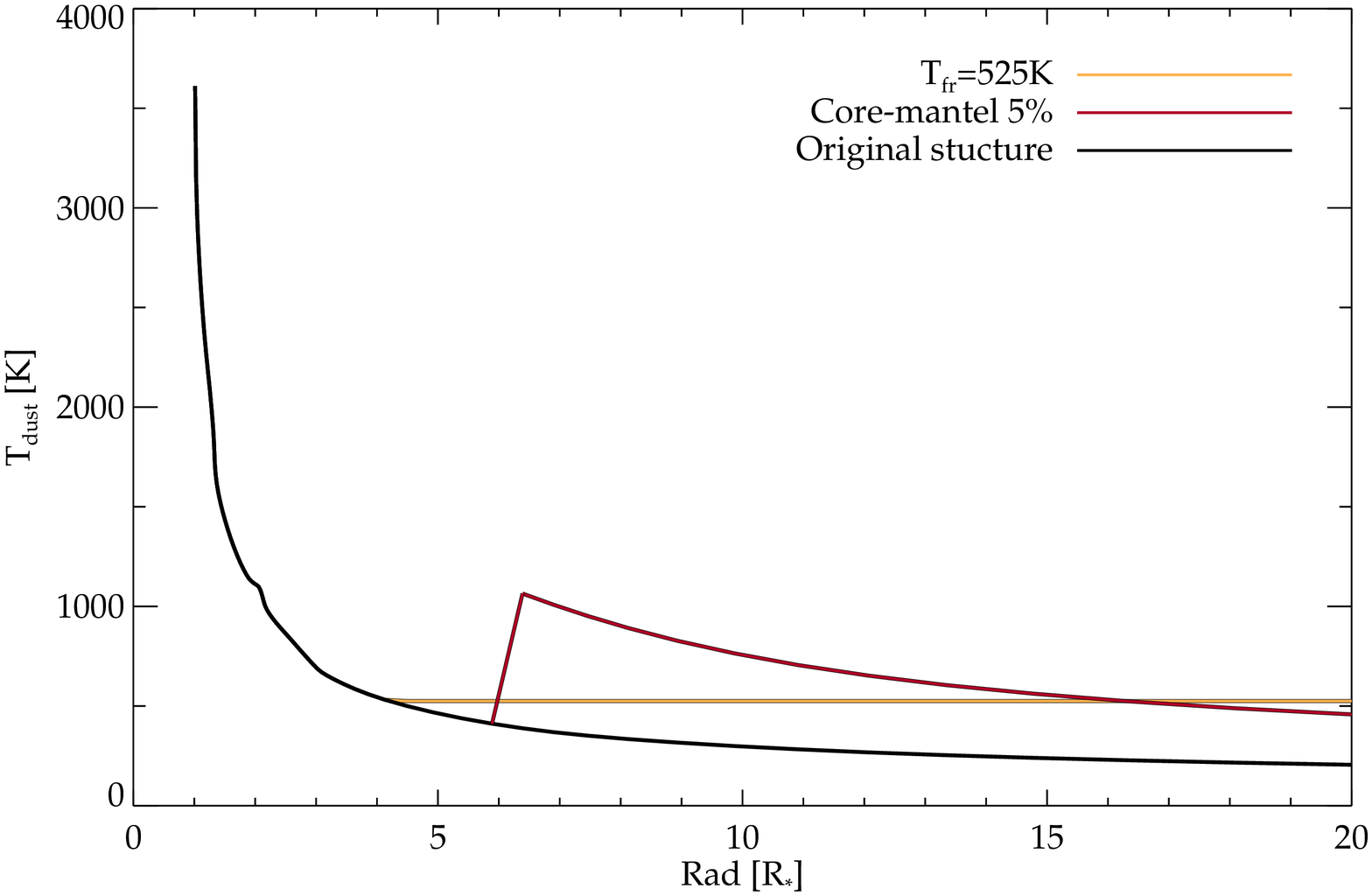}
\includegraphics[width=\linewidth]{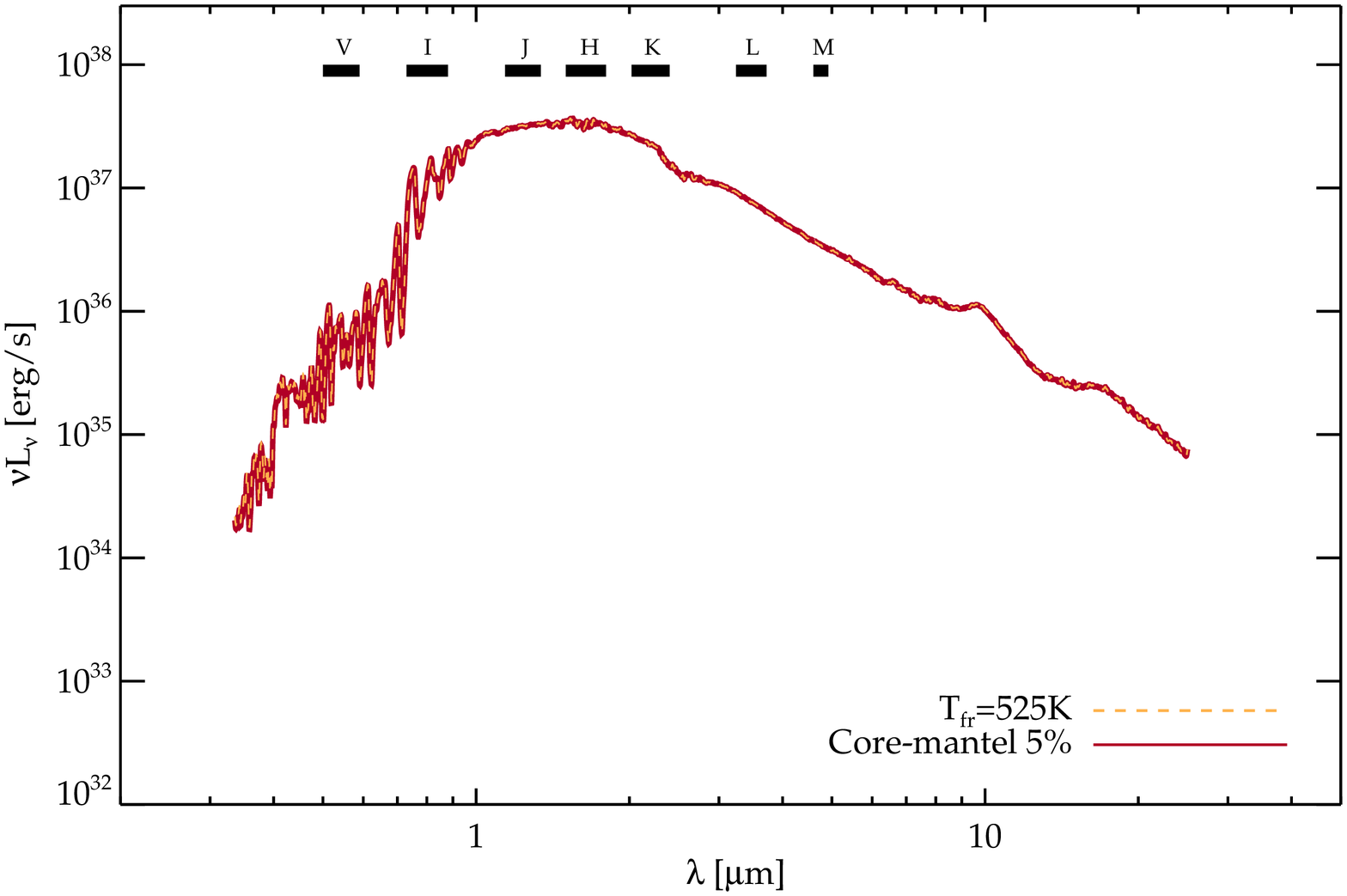}
   \caption{\textit{Top panel:} Grain temperature as a function of distance from the star. \textit{Lower panel:} Spectral energy distributions as a function of wavelength. Structure from model A3 during the maximum luminosity phase. The dark red curves show temperature and spectra from snapshots with core-mantel grains of 5\% thickness. The light red curves show the results from fitting the core-mantel spectra with snapshots of the atmosphere where we set the lower limit for the grain temperature to $525$\,K.}
\label{f_final1}
\end{figure}

\begin{figure}
\centering
\includegraphics[width=\linewidth]{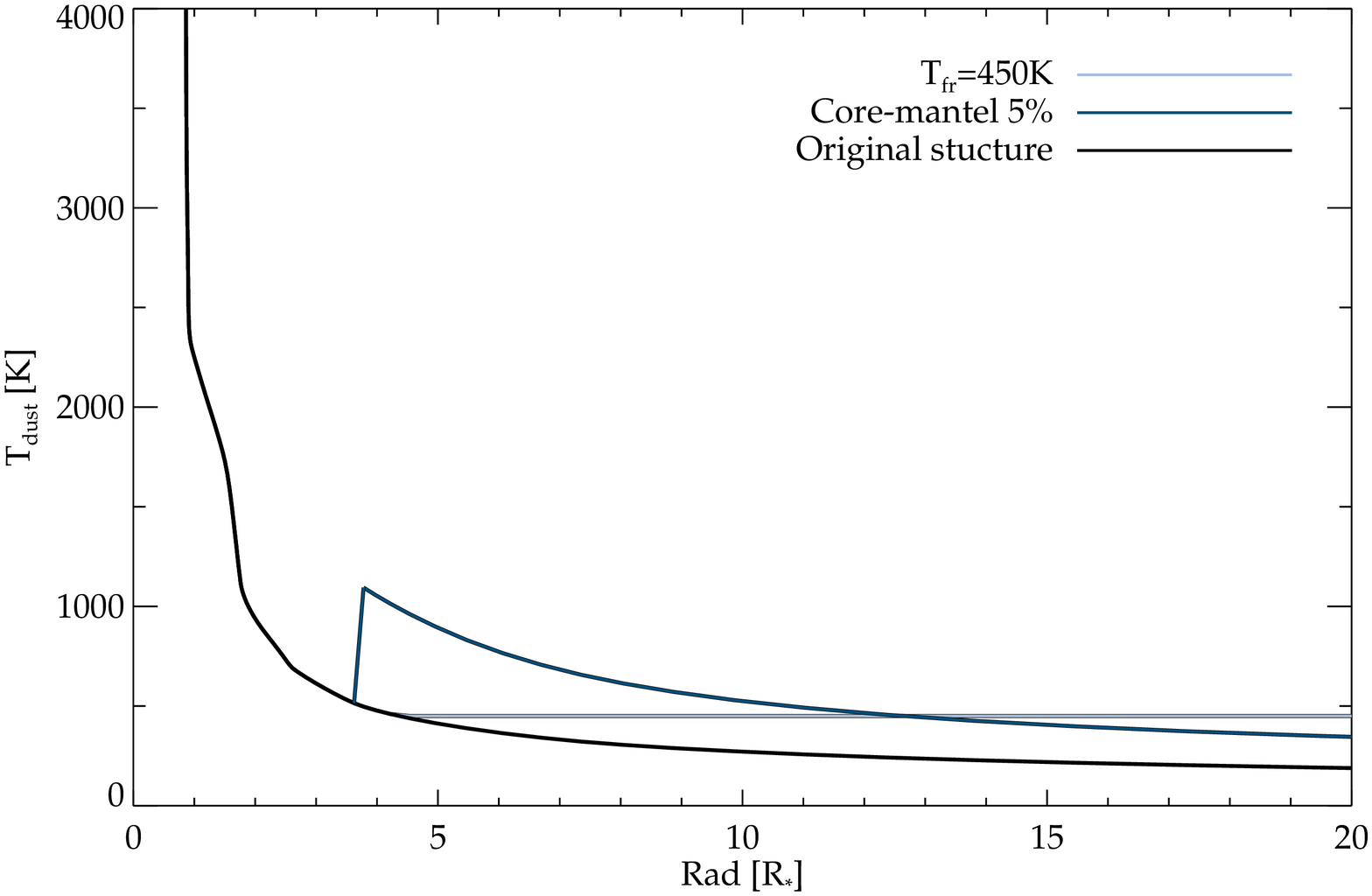}
\includegraphics[width=\linewidth]{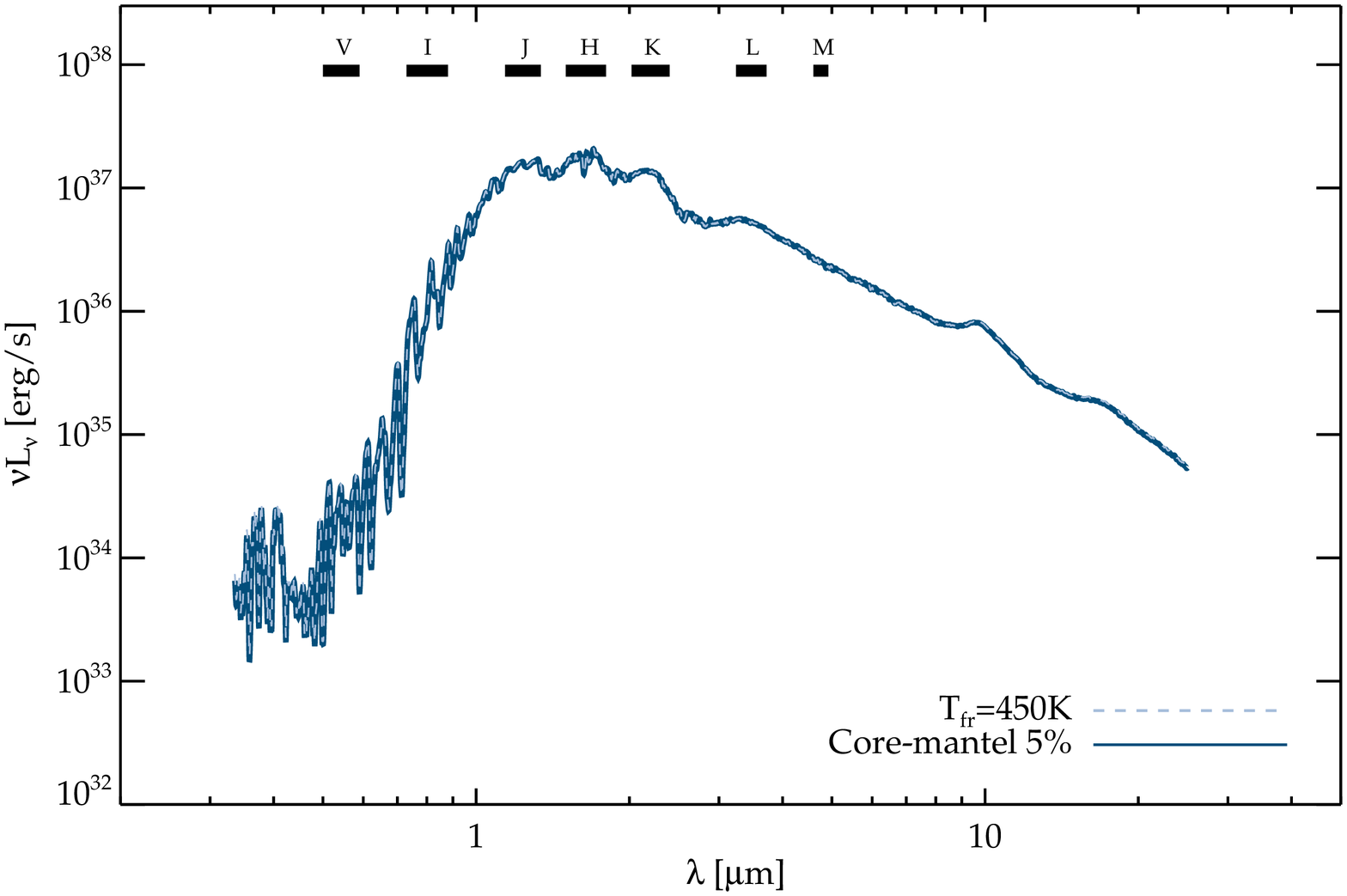}
   \caption{\textit{Top panel:} Grain temperature as a function of distance from the star. \textit{Lower panel:} Spectral energy distributions as a function of wavelength. Structure from model A3 during minimum luminosity phase. The dark blue curves show temperature and spectra from snapshots with core-mantel grains of 5\% thickness. The light blue curves show the results from fitting the core-mantel spectra with snapshots of the atmosphere where we set the lower limit for the grain temperature to $450$\,K.}
    \label{f_final2}
\end{figure}

Given the two strategies presented in Sect.~\ref{gcomp} to study the effects of Fe enrichment in silicates in the DARWIN models for M-type AGB stars, we focus on method (ii), using the spectral energy distributions produced from model A3 with core-mantle grains of 5\% thickness and attempt to reproduced them with spectra resulting from method (i). As can be seen in the bottom panel of Figs.~\ref{f_final1} and \ref{f_final2}, we can produce an almost perfect fit for the 5\% core-mantle spectra during the maximum and the minimum luminosity phase by setting a lower limit to the grain temperature at $T_{\mathrm{fr}}=525$\,K and $T_{\mathrm{fr}}=450$\,K, respectively. It is not possible, however, to produce a spectrum with the same lower limit of the grain temperature that will fit both the maximum and minimum luminosity phase of the core-mantle model. The spectra using a lower limit to the grain temperature are plotted in Figs.~\ref{f_final1} and \ref{f_final2} in a lighter shade under the colored core-mantle spectra. The different temperature structures of the dust component are shown in the top panel of Figs.~\ref{f_final1} and \ref{f_final2}. The radial temperature profiles resulting from the two methods are indeed very different, yet they still produce similar spectra. This shows that the spectral energy distribution alone is not an observable that we can use to differentiate between the temperature profiles or to determine the degree of Fe enrichment as a function of distance.

\subsection{Intensity and visibility profiles}
\label{interfer}
\begin{figure*}
\centering
\includegraphics[width=\linewidth,viewport = 57 544 552 718,clip]{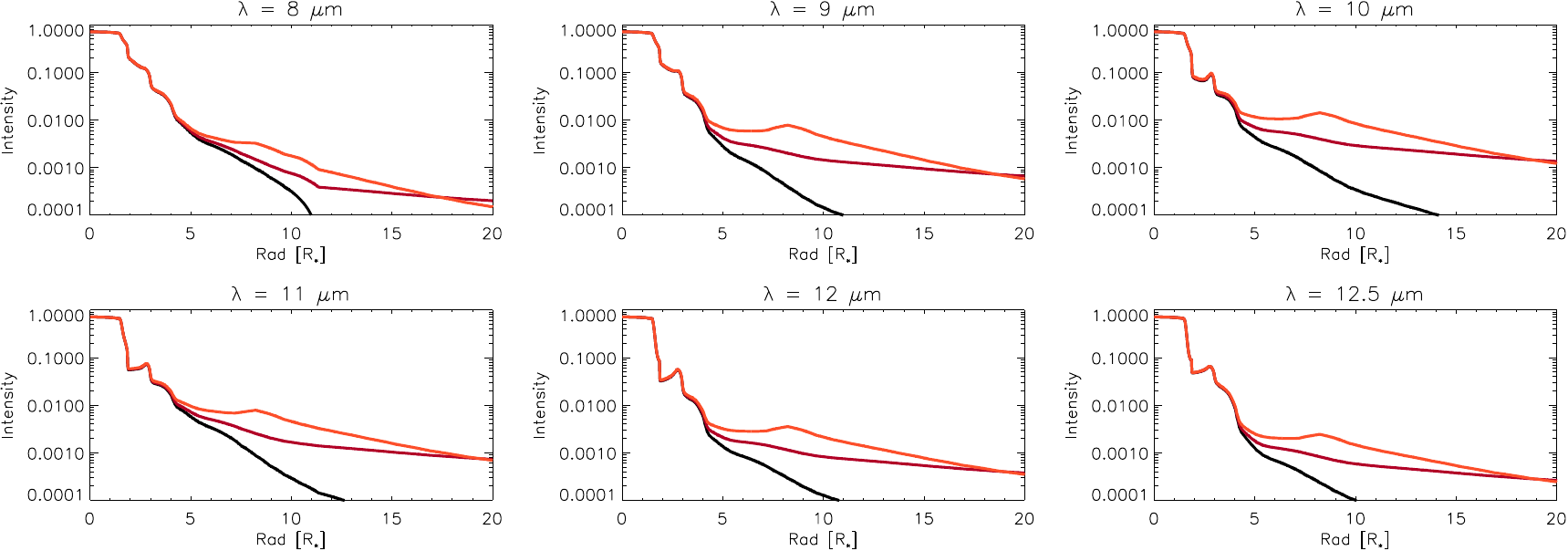}
 \caption{Spatial intensity profile for the models shown in Fig.~\ref{f_final1} (maximum luminosity phase). Every panel corresponds to a wavelength in the N-band interval [8-13 $\mu$m]. The red curve is from a snapshot with core-mantel grains of 5\% thickness. The orange curve show the results from fitting the core-mantel spectra with snapshots of the atmosphere where we set the lower limit for the grain temperature to $525$\,K.}
 \label{intensityfinal1}
\end{figure*}

Figure~\ref{intensityfinal1} shows the spatial intensities at selected wavelengths across the N-band, corresponding to the maximum luminosity models plotted in Fig.~\ref{f_final1}. The intensity profiles of the models are very different from 4 stellar radii on, mirroring the differences introduced in the temperature structures. The N-band intensity distribution of the original model (black line) is more compact than the models where we test different enrichment scenarios. The model with a constant temperature in the outer regions (method (i), orange line) results in brighter profiles than in the other two cases, with a shell-like structure between 8 and 9~R$_*$.

\begin{figure*}
\centering
\includegraphics[width=\linewidth,viewport = 80 369 545 715,clip]{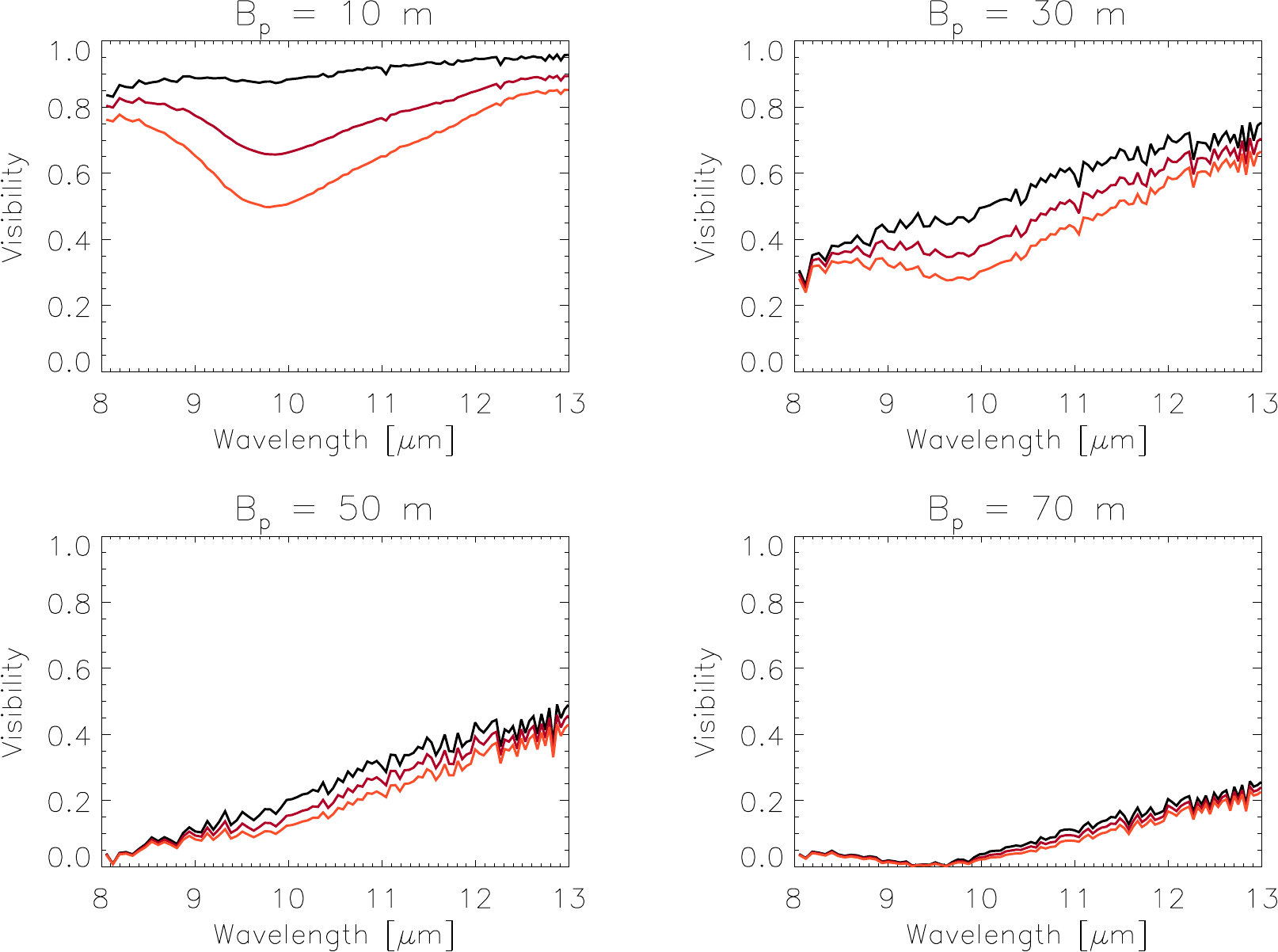}
   \caption{Visibility spectrum across the N band for the models shown in Fig.~\ref{f_final1} (maximum luminosity phase). The panels represent observations simulated with different projected baselines (B$_p = 10, 30, 50,$ and 70 m) regularly available on the VLTI facility. Color-coding as in Fig.~\ref{intensityfinal1}.\label{visibilityfinal1}}
\end{figure*}

The predicted visibility spectrum is very different for the three cases (Fig.~\ref{visibilityfinal1}). The two models with modified temperature structures produce a much more pronounced 10~$\mu$m feature, as can be seen in the predictions for a 10~m projected baseline synthetic visibility. The original model is much flatter than the other two. The differences between the three models should be easily detectable with the data produced with previous (MIDI) and future facilities (MATISSE), as they are larger than a typical 10\% relative error. The intensities and visibilities of the minimum phase show a similar behavior to the one just described for the maximum phase (see Figs.~\ref{intensityfinal_min1} and \ref{visibilityfinal_min1} in the Appendix). 

Observations with baselines that probe spatial scales of about 4 stellar radii and beyond are the most suitable for tracing changes in grain composition, since this is where effects of Fe enrichment should be found. The longer the projected baseline, i.e., the larger the angular resolution of the interferometer, the smaller the differences between the models.
In a very simple (1D) picture, the close environment of the star, where the temperature stratifications of the three cases presented here are the same, is probed at higher angular resolution. However, it should be kept in mind that asymmetric structures play a role at higher angular resolution observations (i.e., visibilities below 0.1), as documented in the literature \citep{deroo2007, ohnaka2008, paladini2012, sacuto2013, paladini2017}. In any investigation including 1D models, such low visibilities should be regarded with care. Nevertheless, it should mentioned here that \cite{sacuto2013} were able to trace silicate signatures in RT Vir down to about 2 stellar radii. According to theoretical models, grains at those distances should be basically Fe-free. In order to put constraints on the wind-driving mechanism it is therefore necessary to cover both the inner wind region (2-4 stellar radii) and the outer layers where significant changes in grain composition can be expected.

\section{Summary and conclusions}
\label{conclusion}
Iron-free magnesium silicate particles with grain sizes of about 0.1 - 1~$\mu$m, which are good candidates for triggering the winds of M-type AGB stars, are too cool to produce the characteristic mid-IR features often observed in these stars \citep[see discussion in][]{bladh2015}. A probable explanation is that silicate grains are gradually enriched with Fe (or other impurities) as they move away from the star, which increases radiative heating.

The aim of this work is to investigate whether this change in grain composition over the first few stellar radii in the outflow can be detected with spatially resolved mid-IR observations. To investigate this we use a detailed model of the atmosphere and wind \citep[produced with the 1D radiation-hydrodynamical code DARWIN, see][]{hoefner2016} and simulate the effects of Fe inclusion in the a posteriori spectral calculations (i) by directly changing the temperature profile of the silicates by setting a lower limit to the grain temperature or (ii) by altering the optical properties of the silicate grains, considering dust particles with a core of pure magnesium silicate (Mg$_2$SiO$_4$) and a thin mantle of magnesium-iron silicate (MgFeSiO$_4$), and then calculating the grain temperature. In both cases the resulting spectra and photometry at visual and near-IR wavelengths ($0.4-5~\mu$m) are largely unaffected, whereas the mid-IR features at 10 and 18$~\mu$m change dramatically (see Figs.~\ref{f_freezing} and \ref{f_coremantel}). This approach assumes that the atmospheric structure will not be altered significantly by the change in grain temperature. This is a reasonable assumption since the dynamical impact on the atmospheric structure is dominated by the radiation pressure generated in the near-IR region, where the spectra remain basically unchanged.

These two methods are then used to construct temperature structures that give rise to essentially the same overall spectral energy distribution (see Figs.~\ref{f_final1} and \ref{f_final2}) but with very different radial profiles. This shows that the spectral energy distribution alone is not an observable that can be used to differentiate between the temperature profiles or to determine the degree of Fe enrichment as a function of distance.

Spatially resolved interferometric observations have the potential to distinguish between the different grain-enrichment scenarios tested here. Observations probing spatial scales larger than about 4 stellar radii are the most suitable ones for this purpose. In order to put constraints on the wind-driving mechanism, however, it is necessary to also cover the inner wind region (2-4 stellar radii) where the silicates should be Fe-free. The next step of this work will be to use wind models to interpret the existing mid-IR spectro-interferometric data.

\begin{acknowledgements}
This work has been supported by the Swedish Research Council (Vetenskapsrådet) and by the ERC Consolidator Grant funding scheme (STARKEY project, G.A. No. 615604). C.P. acknowledges support from the Belgian Fund for Scientific Research F.R.S.-FNRS, and the European Community’s Seventh Framework Programme under Grant Agreement 312430. The computations of spectra and photometry were performed on resources provided by the Swedish National Infrastructure for Computing (SNIC) at UPPMAX.
\end{acknowledgements}

\bibliographystyle{aa}
\bibliography{biblio}
\begin{appendix}
\section{Intensity and visibility profiles}
\begin{figure*}
\centering
\includegraphics[width=\linewidth,viewport = 57 544 552 718,clip]{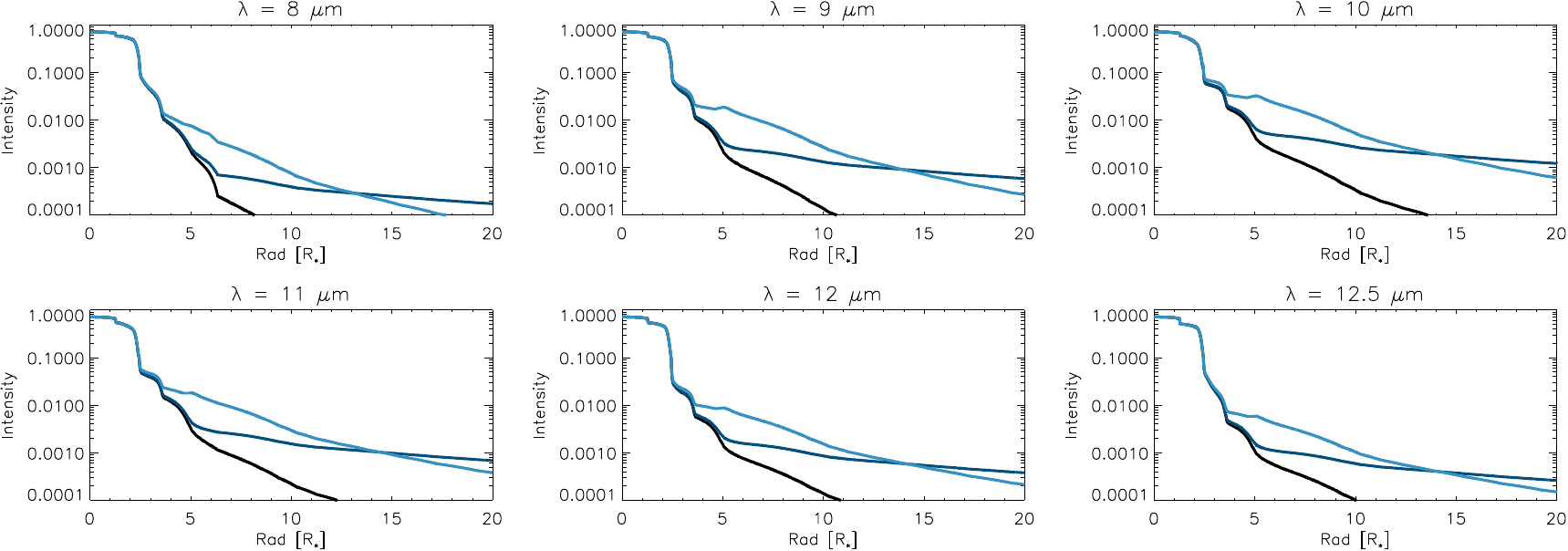}
   \caption{Spatial intensity profile for the models shown in Fig.~\ref{f_final2} (minimum luminosity phase). Each panel corresponds to a wavelength in the N-band interval [8-13 $\mu$m]. The dark blue curve is from a snapshot with core-mantel grains of 5\% thickness. The light blue curve show the results from fitting the core-mantel spectra with snapshots of the atmosphere where we set the lower limit for the grain temperature to $450$\,K.}
    \label{intensityfinal_min1}
\end{figure*}

\begin{figure*}
\centering
\includegraphics[width=\linewidth,viewport = 80 369 545 715,clip]{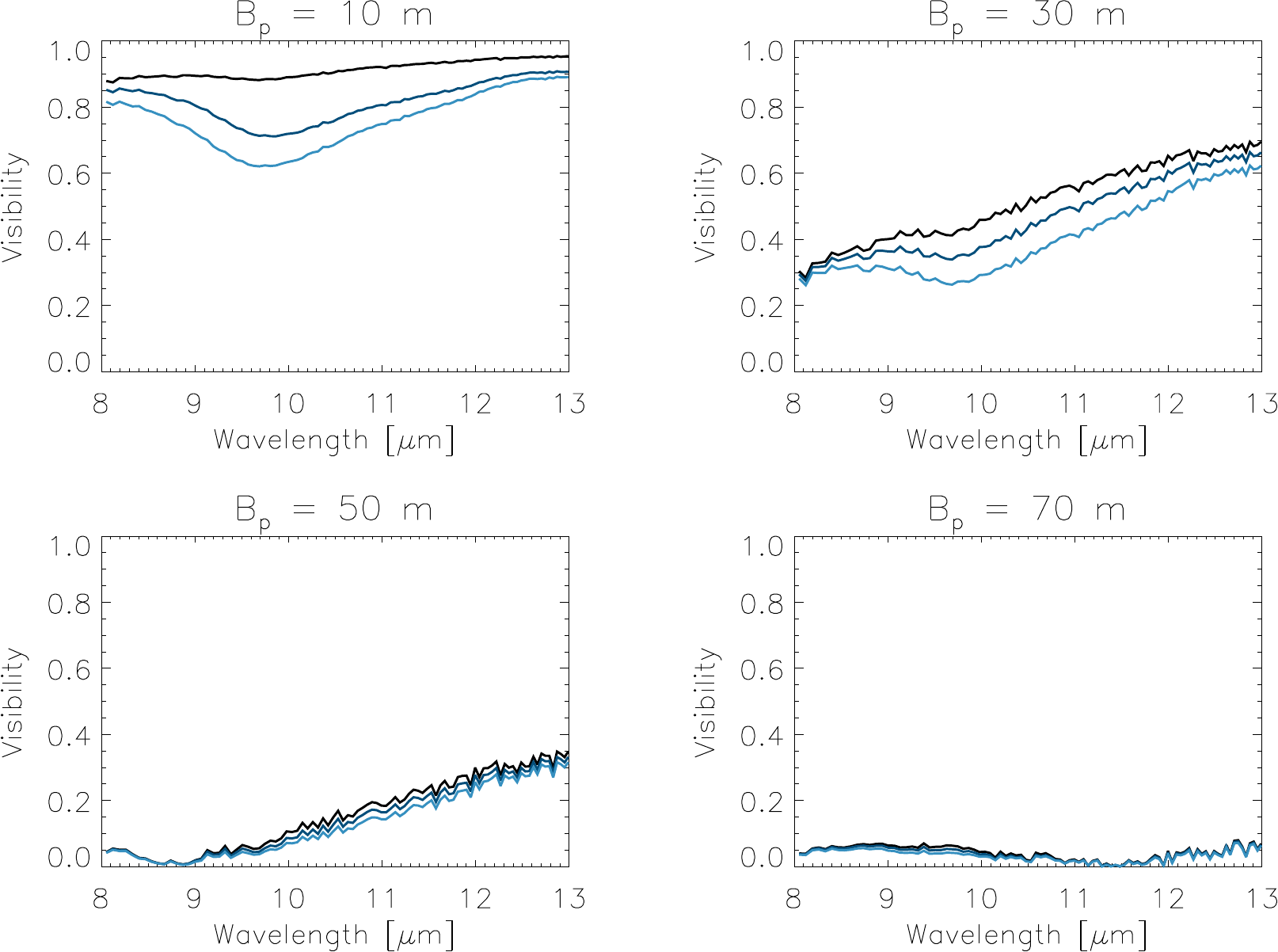}
   \caption{Visibility spectrum across the N band for the models shown in Fig.~\ref{f_final1} (minimum luminosity phase). The panels represent observations simulated with different projected baselines (B$_p = 10, 30, 50,$ and 70 m) regularly available on the VLTI facility. Color-coding as in Fig.~\ref{intensityfinal_min1}.}
    \label{visibilityfinal_min1}
\end{figure*}

\begin{figure*}
\centering
\includegraphics[width=\linewidth,viewport = 67 484 554 718,clip]{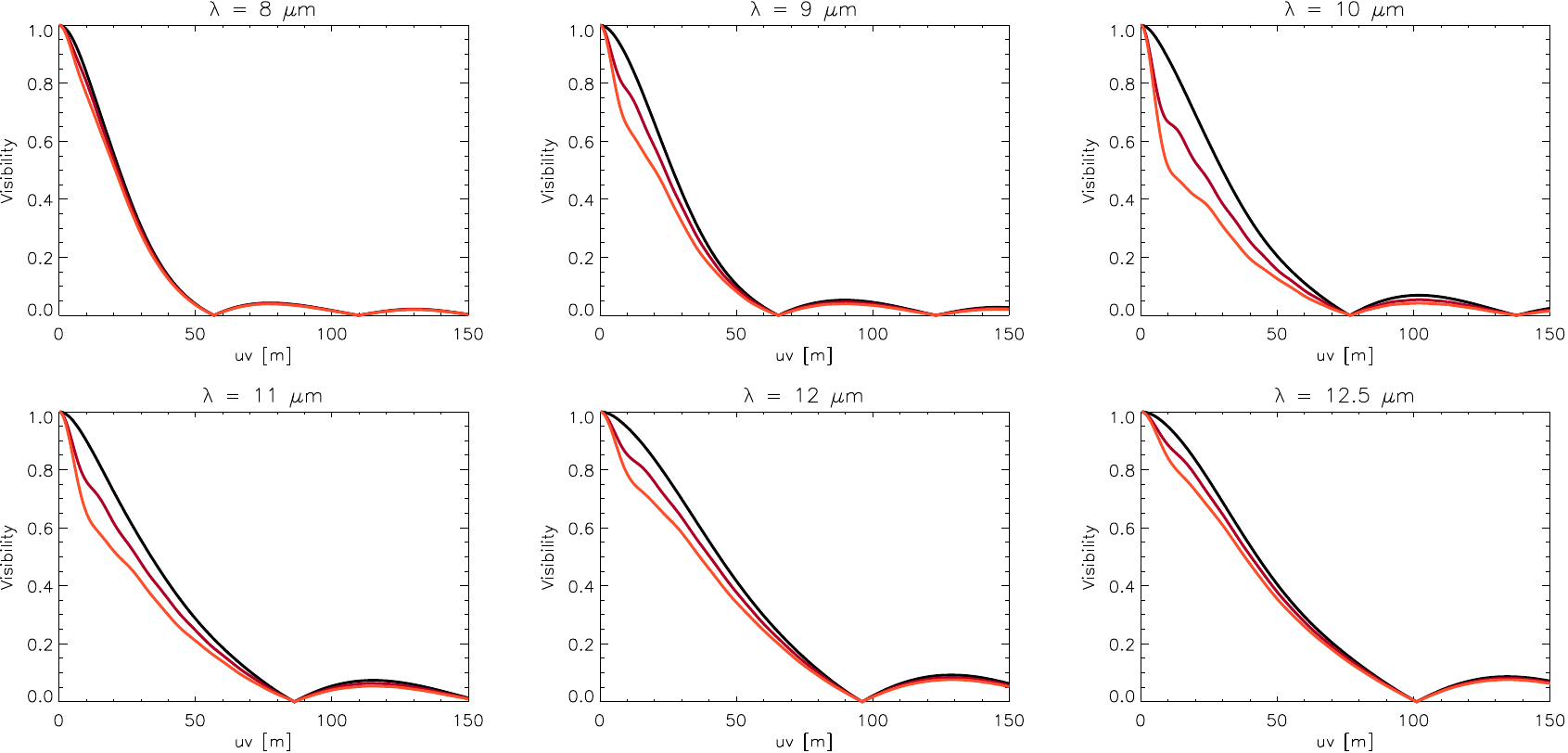}
   \caption{Visibility versus spatial frequency for the models shown in Fig.~\ref{f_final1} (maximum luminosity phase). The various panels correspond to different wavelengths across the N band. Color-coding as in Fig.~\ref{intensityfinal1}.}
   \label{visibility_final1}
\end{figure*}

\begin{figure*}
\centering
\includegraphics[width=\linewidth,viewport = 67 484 554 718,clip]{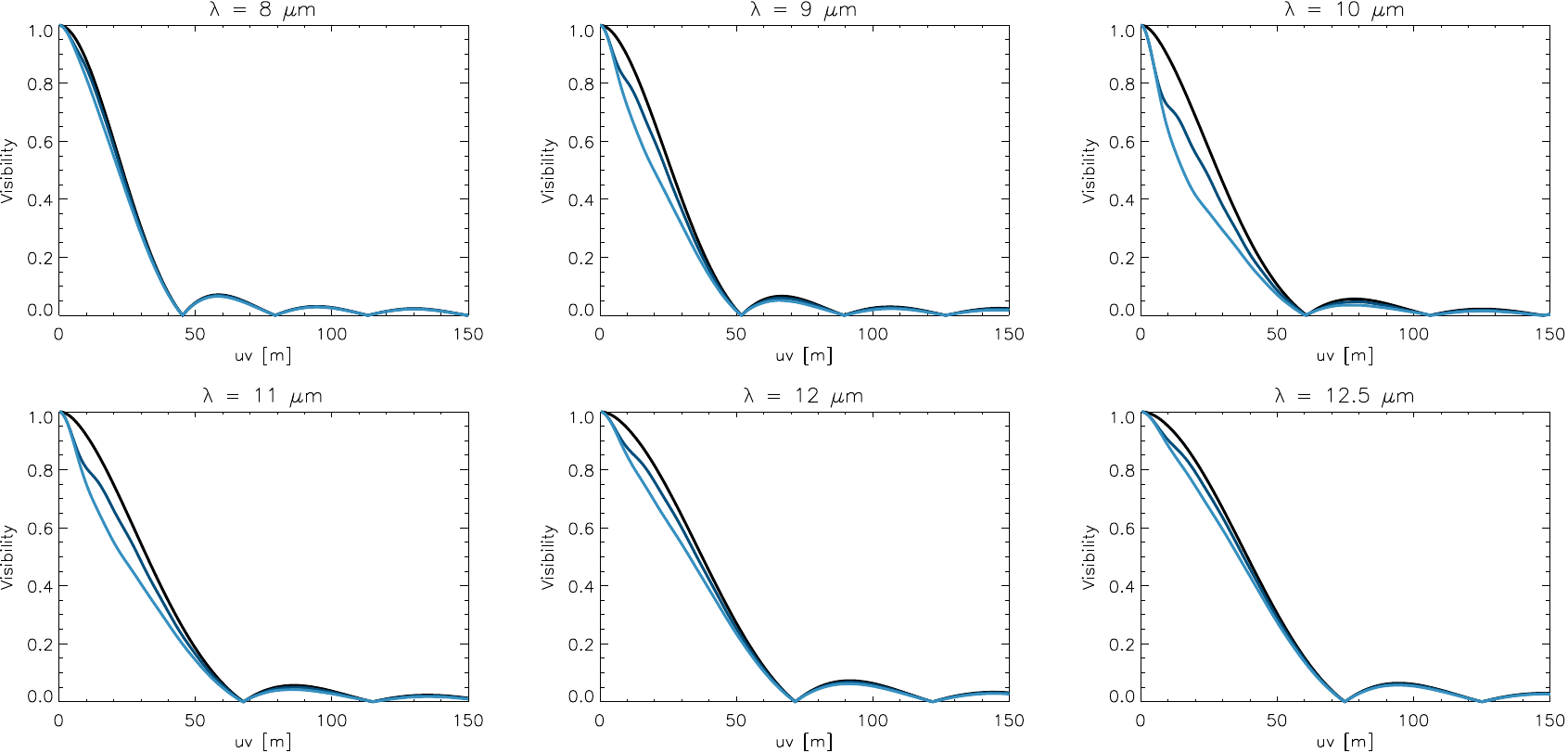}
   \caption{Visibility versus spatial frequency for the models shown in Fig.~\ref{f_final2} (minimum luminosity phase). The various panels correspond to different wavelengths across the N band. Color-coding as in Fig.~\ref{intensityfinal_min1}}
    \label{visibility_final2}
\end{figure*}

Synthetic intensity (Fig.~\ref{intensityfinal_min1}) and visibility profiles (Fig.~\ref{visibilityfinal_min1}) for the minimum luminosity phase, and visibility profile predictions versus spatial frequencies for the maximum (Fig.~\ref{visibility_final1}) and the minimum (Fig.~\ref{visibility_final2}) luminosity phase.

\end{appendix} 

\end{document}